\documentstyle[12pt,aasms4]{article}
\newcommand{\postscript}[2]
 {\setlength{\epsfxsize}{#2\hsize}
   \centerline{\epsfbox{#1}}}

\begin{document}

\title{Unique Determination of the Physical Parameters of \\
       Individual MACHOs from Astrometric Parallax Measurements}

\bigskip
\bigskip
%---authors
\author{Cheongho Han}
\affil{Department of Astronomy \& Space Science, \\
Chungbuk National University, Chongju, Korea 361-763 \\
cheongho@astronomy.chungbuk.ac.kr,}
\authoremail{cheongho@astronomy.chungbuk.ac.kr}

%\bigskip
%\centerline{\it \&}
%\bigskip
%\author{Kyongae Chang}
%\affil{Department of Physics, \\
%       Chongju University, Chongju, Korea 360-764 \\
%       kchang@alpha94.chongju.ac.kr}
% \authoremail{kchang@alpha94.chongju.ac.kr}

\bigskip
\bigskip

%======= ABSTRACT =================================================
\begin{abstract}
The Einstein time scale, which is the only information obtained from current 
microlensing experiments, results from a complicated combination of 
the lens parameters that we want to determine.  Of the methods for 
breaking the lens parameter degeneracy, the most promising and generally 
applicable method is to measure the lens parallax from simultaneous 
observations of a lensing event from the ground and a heliocentric 
satellite (Gould 1994).  However, the elegant idea of parallax measurement, 
which was proposed to resolve the lens parameter degeneracy, paradoxically 
suffers from its own degeneracy due to the ambiguity of the source 
star trajectory.  In this paper, we propose to measure the lens 
parallax astrometrically by mounting an interferometer instead of a 
photometer in the proposed parallax satellite.  By simultaneously 
measuring the source star centroid shifts from a geocentric and an 
additional heliocentric satellite, one can determine the lens parallax 
without ambiguity.  If the already planned {\it Space Interferometry 
Mission} will be used as one of the satellites, one simply needs to  
replace the photometer in the parallax satellite with an instrument 
for astrometric observation.  In addition, since the proposed method 
can measure both the lens parallax and the proper motion at the same time,
one can completely break the lens parameter degeneracy, and therefore 
uniquely determine the physical parameters of individual lenses.

\end{abstract}

\vskip20mm

\centerline{submitted to {\it Monthly Notices of the 
Royal Astronomical Society}: Feb 5, 1999}
\centerline{Preprint: CNU-A\&SS-02/99}
\clearpage

%==== Main Body ==================================================
\section{Introduction}
Searches for gravitational microlensing events caused by Massive 
Astronomical Compact Objects (MACHOs) by monitoring stars 
in the Large Magellanic Cloud (LMC) and the Galactic bulge are currently 
underway and several hundred events have been detected to date (Alcock et 
al.\ 1997a, 1997b; Ansari et al.\ 1996; Udalski et al.\ 1997; Alard \& 
Guibert 1997).\markcite{alcock1997, ansari1996, udalski1997, alard1997}  
The only information about the lens obtained from current lensing experiments
is the Einstein time scale.  However, the time scale results from a 
complicated combination of the physical parameters of the lens by
$$
t_{\rm E} = {r_{\rm E} \over v}; \qquad
r_{\rm E} = \left( {4GM\over c^2}{D_{ol}D_{ls}\over D_{os}}\right)^{1/2},
\eqno(1.1)
$$
where $r_{\rm E}$ is the physical size of the Einstein ring, $M$ is the
mass of the lens, $v$ is the lens-source transverse speed, and $D_{ol}$,
$D_{ls}$, and $D_{os}$ are the separations between the observer, lens,
and source star.  Due to the lens parameter degeneracy, our 
knowledge about the nature of MACHOs is very poor in spite of the large number 
of detected events.

Numerous methods for resolving the lens parameter degeneracy have been 
proposed. While most of these methods have very limited applications, two 
methods are applicable to microlensing events in general. The first method is 
to measure the lens parallax by simultaneously observing a lensing event from 
the ground and a heliocentric satellite (Gould 1994, 1995).  If the parallax 
of an event is measured, one can determine the projected speed 
$\tilde{v}=(D_{ol}/D_{ls})v$ and the lens parameter degeneracy can be
partially resolved (see \S\ 2).  The second method is to measure the lens 
proper motion, $\mu$, by astrometrically measuring the source star 
centroid shifts caused by gravitational lensing with a high precision 
interferometer such as the {\it Space Interferometry Mission} (hereafter 
SIM, http://sim.jpl.nasa.gov/).  When a lensing event is astrometrically 
observed, one can determine the lens proper motion, which can also 
partially resolve the lens parameter degeneracy (Miyamoto \& Yoshii 1995; 
Walker 1995; H\o\hskip-1pt g, Novikov \& Polarev 1995; 
Boden, Shao, \& Van Buren 1998).  When both $\tilde{v}$ and $\mu$ are 
determined, the degeneracy is completely broken and the lens parameters 
of individual lenses are determined by
$$
\cases{
M = (c^2/4G)t_{\rm E}^2\tilde{v}\mu,\cr
D_{ol} = D_{os}(\mu D_{os}/\tilde{v} +1)^{-1},\cr
v = [\tilde{v}^{-1}+(\mu D_{os})^{-1}]^{-1}.
}
\eqno(1.2)
$$

Unfortunately, the elegant idea of lens parallax measurements, which was 
proposed to resolve the lens parameter degeneracy, paradoxically suffers 
from its own degeneracy.  This degeneracy from the parallax measurement 
arises because one cannot uniquely determine the source star trajectory 
from the photometrically constructed light curve alone.  As a result, the 
measured projected speed suffers from a two-fold degeneracy (see \S\ 2).

In this paper, we propose to measure the lens parallax astrometrically
by mounting an interferometer instead of a photometer in the proposed 
parallax satellite.  By simultaneously measuring the source star centroid 
shifts from the SIM and the astrometric parallax mission, one can determine 
the lens parallax without any ambiguity.  In addition, since the proposed 
method can measure both the lens parallax and proper motion at the same time, 
one can completely break the lens parameter degeneracy, and thus the 
physical parameters of individual lenses can be uniquely determined.

\section{Degeneracy in the Projected Velocity}
When seen from a satellite with a two-dimensional separation vector
${\bf r}$ with respect to the Earth, the Einstein ring will be displaced by
an angle
$$
\Delta\vec{\theta} = (D_{os}^{-1}-D_{ol}^{-1}){\bf r} 
= -{D_{ls}\over D_{os}}{{\bf r}\over D_{ol}}
\eqno(2.1)
$$
relative to its position as seen from the Earth.
Due to the displacement of the Einstein ring, the source star 
position with respect to the lens is displaced by the
same amount $\Delta\vec{\theta}$, but towards the opposite direction.
When normalized by the Einstein ring radius, the displacement vector of
the source star position (i.e.\ parallax)
$\Delta{\bf u} = (\Delta t_0/t_{\rm E})\ {\bf x}' + \beta\ {\bf y}'$
is then related to the Earth-satellite separation vector by
$$
\Delta{\bf u} = -{D_{ol}\Delta\vec{\theta}\over r_{\rm E}}
= {{\bf r}\over (D_{os}/D_{ls})r_{\rm E}}.
\eqno(2.2)
$$
By multiplying $t_{\rm E}$ by both sides of equation (2.2) and using the
definition of the projected velocity of $\tilde{\bf v}=(D_{os}/D_{ls}){\bf v}$,
one finds the relation between the projected velocity and the parallax by
$$
\tilde{\bf v} = \tilde{v}_{x'}\ \hat{\bf x'} + 
\tilde{v}_{y'}\ \hat{\bf y'},
\eqno(2.3)
$$
where
$$
\tilde{v}_{x'} = {r\over t_{\rm E}\Delta u^2} {\Delta t_0\over t_{\rm E}}
\eqno(2.4)
$$
and
$$
\tilde{v}_{y'} = {r\over t_{\rm E}\Delta u^2} \Delta\beta.
\eqno(2.5)
$$
Here $\hat{\bf x'}$ and $\hat{\bf y'}$ are the unit vectors which are 
parallel and perpendicular to the Earth-satellite separation vector and 
$\Delta t_0$ and $\Delta\beta$ are the differences in the times of maximum 
amplification and the impact parameters of the two light curves, respectively. 
Since the event light curve is related to the lensing parameters by
$$
A = {u^2+2 \over u(u^2+4)^{1/2}};\qquad 
u^2 = \left({t-t_0\over t_{\rm E}} \right)^2 + \beta^2,
\eqno(2.6)
$$
the values of $t_0$ and $\beta$ are determined from 
the individual light curves.  Therefore, by determining 
$\Delta u = [(\Delta t_0/t_{\rm E})^2 + \Delta\beta]^{1/2}$ 
one can determine the transverse speed by
$$
\tilde{v} = {r\over t_{\rm E}} {1\over \Delta u}.
\eqno(2.7)
$$

However, the photometrically determined parallax suffers from two-fold 
degeneracy.  This degeneracy occurs because events with different 
trajectories can have the same lensing parameters, resulting in the same 
light curve.  In Figure 1, we illustrate this degeneracy in the photometric 
parallax measurements.  The upper panel shows the light curves seen from the
Earth and the satellite, respectively.  Since the shift in the Einstein ring
does not change its size, the two light curves have the same Einstein
time scale.  On the other hand, the time of maximum amplification and
the impact parameter are changed due to the shift of the Einstein ring,
resulting in different light curves (Refsdal 1966).  In the middle panels, 
we present the two possible lens system geometries which can produce the 
light curves in the upper panel.  The source star trajectories as seen 
from the Earth and satellite are represented by solid lines.  The circle 
(represented by dashed lines) around the lens ({\it L}) represents the 
Einstein ring.  The coordinates $(x,y)$ are chosen so that they are 
parallel and perpendicular to the source star trajectory and they are 
centered at the lens position.  The bold vectors $\Delta {\bf u}$ 
connecting two points ($S_{\rm s}$ for the source seen from the satellite 
and $S_{\rm E}$ from the Earth) on individual trajectories represent the 
displacements of the source star position (i.e.\ parallax) observed 
at a moment.  Note that the amount of parallax has a different value 
depending on the lens system geometry.  Since the projected speed is 
inversely proportional to $\Delta u$ [see equation (2.6)], there will 
be two possible values of $\tilde{v}$ (Gould 1994).

\section{Astrometric Source Star Centroid Shifts}
When a source star is gravitationally lensed, it is split into two
images located on the same and opposite sides of the lens, respectively.
Due to the changes in position and amplification of the individual images
caused by the lens-source transverse motion, the light centroid between
the images changes its location during the event.  The location of the image
centroid relative to the source star is related to the lensing parameters by
$$
\delta\vec{\theta}_{c} = {\theta_{\rm E}\over u^2+2}
\left( {t-t_0\over t_{\rm E}}\ \hat{\bf x}+\beta\ \hat{\bf y}\right),
\eqno(3.1)
$$
where $\hat{\bf x}$ and $\hat{\bf y}$ are the unit vectors toward the 
directions which are parallel and normal to the lens-source transverse 
motion and $\theta_{\rm E}=r_{\rm E}/D_{ol}$ is the angular Einstein ring 
radius.  If we let $(x,y)=(\delta\theta_{c,x},\delta\theta_{c,y}-b)$ and
$b=\beta\theta_{\rm E}/2(\beta^2 +2)^{1/2}$, the coordinates are related by
$$
x^2 + {y^2\over q^2} = a^2,
\eqno(3.2)
$$
where $a=\theta_{\rm E}/2(\beta^2+2)^{1/2}$ and 
$q=b/a=\beta/(\beta^2+2)^{1/2}$.  Therefore, during the event the trajectory 
of the source star image centroid traces an ellipse (`astrometric ellipse', 
Walker 1995; Jeong, Han, \& Park 1999).  With the measured astrometric 
ellipse, one can determine the impact parameter of the event because the 
shape (i.e.\ the axis ratio) of the ellipse is related to the impact 
parameter.  In addition, one can determine the Einstein ring radius because 
the size of the astrometric ellipse (i.e.\ semi-major axis) is directly 
proportional to $\theta_{\rm E}$.  While the Einstein time scale, which is 
the only measurable quantity from the event light curve, depends on three 
lens parameters ($M$, $D_{ol}$, and $v$), the angular Einstein ring radius 
depends only on two parameters ($M$ and $D_{ol}$).  Therefore, by measuring 
$\theta_{\rm E}$, the uncertainty in the lens parameter can be significantly 
reduced (Paczy\'nski 1998; Boden et al.\ 1998; Han \& Chang 1999).  Once 
$\theta_{\rm E}$ is determined, the lens proper motion is determined by 
$\mu = \theta_{\rm E}/t_{\rm E}$.

\section{Astrometric Parallax Measurements}
In addition to allowing one to determine lens proper motions, astrometric 
observations of lensing events allow one to uniquely determine source 
star trajectories.  In Figure 2, we present various source star 
trajectories and their corresponding astrometric centroid shifts.  
To distinguish different trajectories, we define an approaching angle 
$\phi$ by the angle between the vector connecting the lens and the 
source at its closest approach and an arbitrary reference direction 
(e.g.\ north).  Then each trajectory is defined by its impact parameter
and the approaching angle.  We additionally define the sign of the 
impact parameter by
$$
{\rm sign}\ (\beta) = 
\cases{
+ , & when $-\pi/2\leq\phi < \pi/2$ \cr
- , & when $\pi/2\leq\phi < 3\pi/2$ \cr},
\eqno(4.1)
$$
and it is marked on each trajectory along with its impact parameter in 
the figure.  One finds that the orientation of the astrometric ellipse, 
which is measured by the angle between the reference direction and the 
semi-minor axis of the ellipse, is identical to the approaching angle of the 
source star trajectory.  Note that the semi-minor axis of the astrometric 
ellipse is identical to the centroid shift at maximum amplification.
Therefore, the two trajectories with the same impact parameter but
with opposite signs, which caused the degeneracy in the photometrically 
determined parallax (i.e. $\beta$ and $-\beta$), result in
astrometric ellipses with opposite orientations.

Since the source star trajectory is uniquely determined from the 
astrometric observations of a lensing event, if the lens parallax is measured 
astrometrically instead via the photometric method, the degeneracy in 
$\Delta{\bf u}$ can be broken, and thus one can uniquely determine 
$\tilde{v}$.  In the lower panel of Figure 1, we present the two sets 
of the source star centroid shifts as seen from the Earth and satellite 
which are expected from the corresponding two sets of source star 
trajectories in the middle panels.  One finds that these two sets of 
astrometric ellipses can be easily distinguished from one another.

\clearpage
\section{Discussion}
Besides the proposed astrometric parallax measurements, the degeneracy 
of the projected speed can also be broken by other methods.  
First, one can, in principle, break the degeneracy in $\tilde{v}$
by measuring the fractional difference in inverse time scales
$\omega=1/t_{\rm E}$ between the Earth and satellite, $\Delta\omega/\omega$,
caused by the relative motion of the Earth and satellite, $v_{s}$.
The larger the projected speed of the lens relative to $v_{s}$, the smaller
$\Delta\omega/\omega$.  Hence the time scale differences allows one to
choose the correct solution (Gould 1996; Boutreux \& Gould 1996;
Gaudi \& Gould 1997).  However, the uncertainty in 
$\tilde{v}$ using this method would be very large.  This is because
the expected time scale difference is very small due to the small relative
velocity between the Earth and the satellite.  To measure the small
difference in time scale, therefore, one must construct light
curves with very high photometric precision.  In addition, both the
Galactic bulge and LMC fields are very crowded, and thus nearly all events
suffer from severe blending.  One might construct a light curve which is
free from blending from diffraction-limited observations from the
satellite (Han 1997).  However, even with very high precision photometric 
observations from the ground, it would be very difficult to determine the time 
scale with an uncertainty small enough to determine $\Delta \omega$
(Wo\'zniak \& Paczy\'nski 1997; Han \& Kim 1999).

Secondly, one can also resolve the degeneracy in $\tilde{v}$ by 
observing a lensing event from a second heliocentric 
satellite located at a different position.  Comparison of the light curves 
observed from the Earth and the second satellite would yield another 
two values for $\tilde{v}$.  Among these values, only one would agree 
well with one of the values of $\tilde{v}$ determined from the first 
satellite parallax measurements, allowing one to select the right 
solution (Gould 1994).  The problem with this method is that it 
requires a very costly second heliocentric satellite.

Of course, the proposed astrometric parallax method also requires
two satellites; one geocentric and the other heliocentric satellite.
However, the SIM is planned for launch regardless of our
proposal.  If the SIM will be used as one of the satellites for
astrometric parallax measurements, then what is required is
simply replacing the photometer in the already proposed photometric
parallax satellite with instruments for astrometric observation.
In addition, since the required astrometric instruments are being developed 
for the SIM mission, by adopting the same instrument one can minimize 
the development cost, which is a significant portion of the total cost.
Finally, and most importantly, by determining both the parallax and 
proper motion at the same time, one can completely resolve the 
lens parameter degeneracy and thus uniquely determine the physical 
parameters of individual lenses.

\acknowledgements
We learned that very similar work to this was independently in progress by 
A.\ Gould and S.\ Salim after most of our work was completed.
We would like to thank P.\ Martini for a careful reading of the manuscript.

\clearpage

\postscript{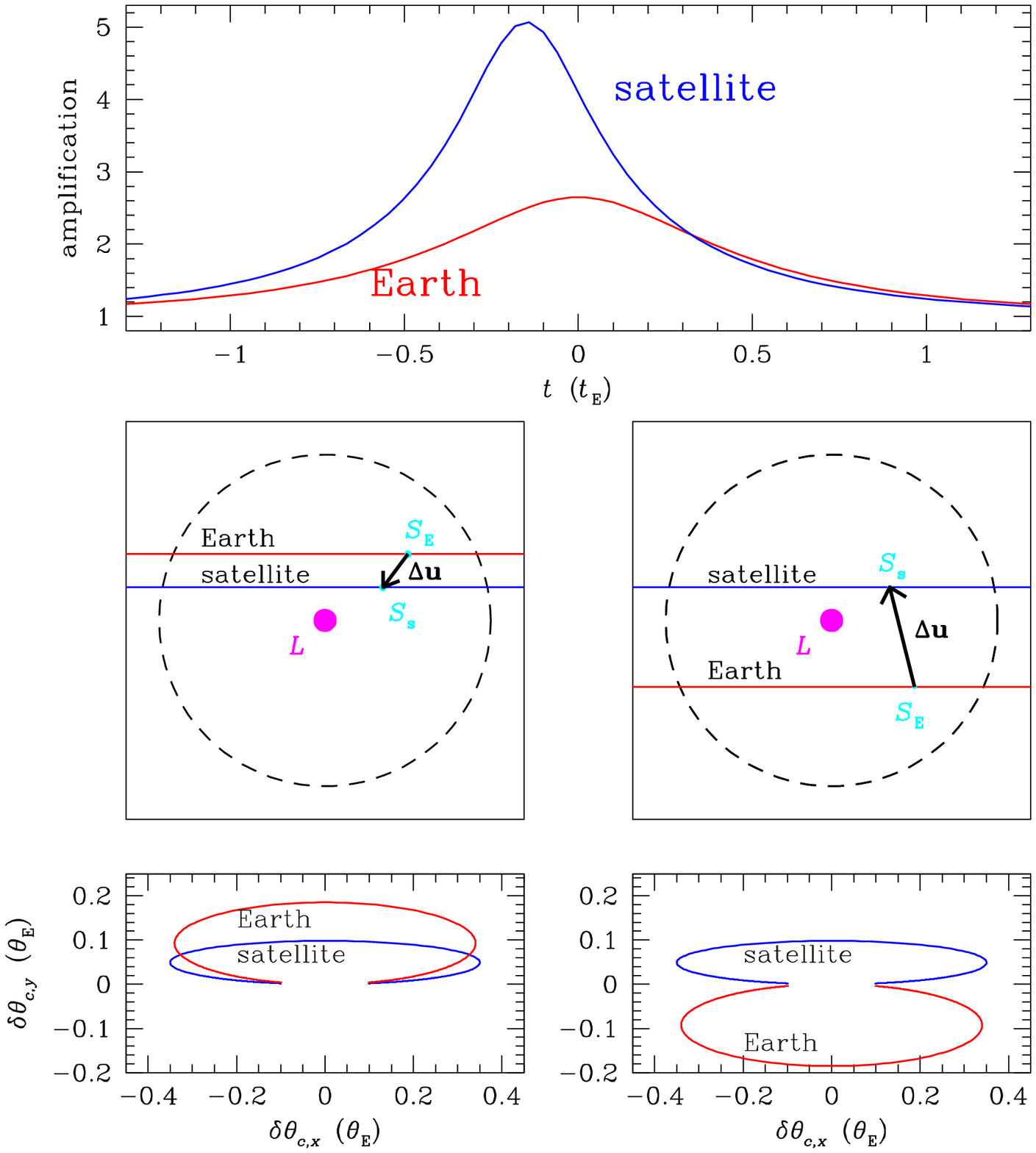}{0.87}
\noindent
{\footnotesize {\bf Figure 1:}\ 
The degeneracy in the photometric parallax measurements.  The upper panel 
shows the light curves seen from the Earth and satellite.  In the 
middle panels are the two possible lens system geometries which produce the 
light curves in the upper panel.  The source star trajectories as seen 
from the Earth and the satellite are represented by solid lines.  The circles 
(represented by dashed lines) around the lenses ({\it L})  are the Einstein 
rings.  The coordinates $(x,y)$ are chosen so that they are parallel and 
perpendicular to the source star trajectory and centered at the lens 
position.  The bold vectors $\Delta {\bf u}$ connecting the two points 
($S_{\rm s}$ for the source seen from the satellite and $S_{\rm E}$ from 
the Earth) on individual trajectories represent the displacements of the 
source star positions (i.e.\ parallaxes) observed at a given time.  Note that 
the parallax has a different value depending on the geometry.  In the 
lower panels, we present the two sets of source star centroid shifts as seen 
from the Earth (geocentric satellite) and from the (heliocentric) satellite 
corresponding to the source star trajectories in the middle panels.
} \clearpage

\postscript{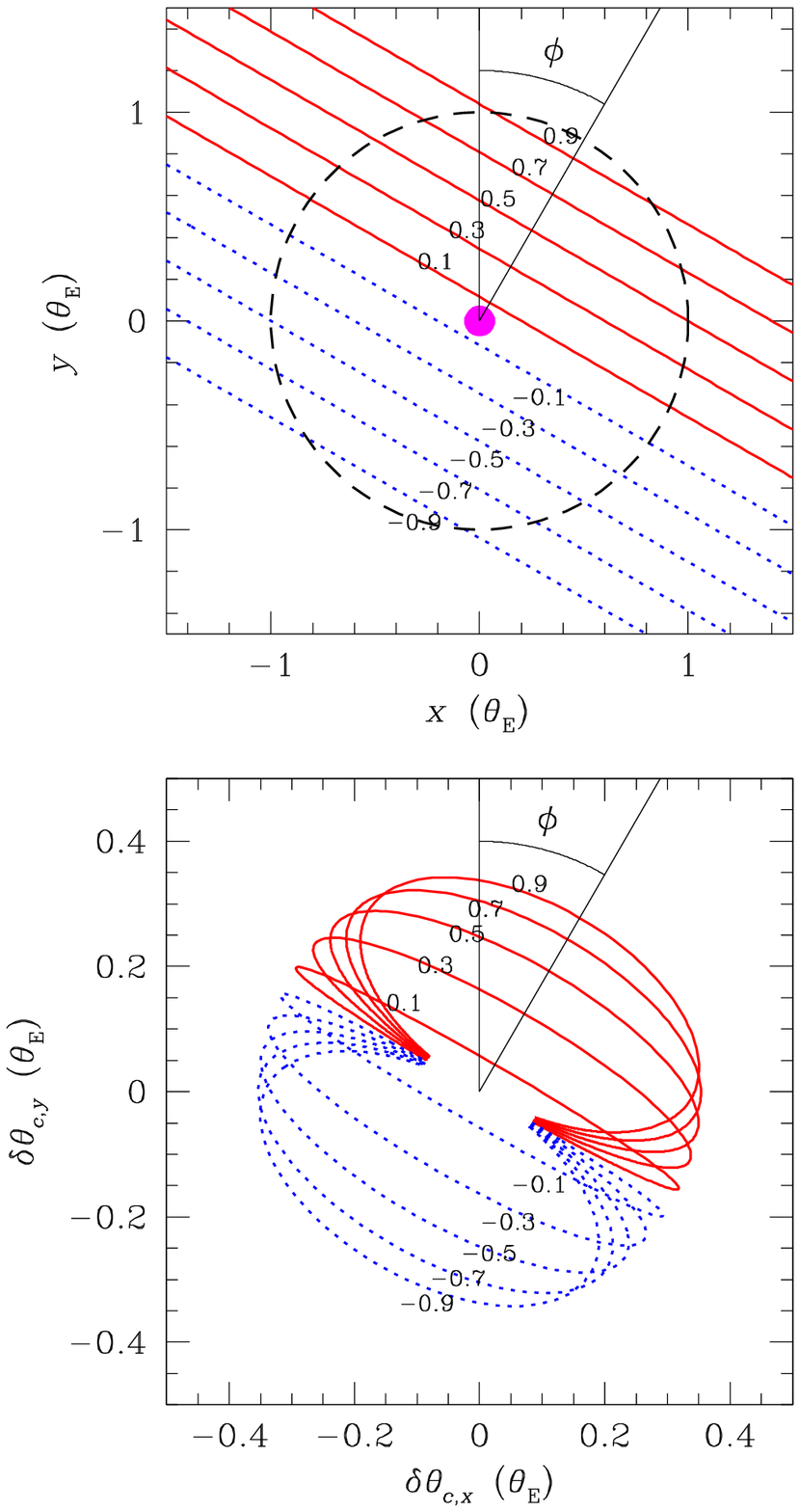}{1.0}
\noindent
{\footnotesize {\bf Figure 2:}\ 
Various source star trajectories and their corresponding astrometric centroid
shifts.  The numbers are the impact parameters of the individual
trajectories.  The approaching angle $\phi$ and the sign of $\beta$ are
defined in the text.  
One finds that the orientation of the astrometric ellipse, which is 
measured by the angle between the reference direction (e.g.\ north)
and the semi-minor axis of the ellipse, is identical to the approaching
angle of the source star trajectory.
Note that the semi-minor axis of the astrometric ellipse is identical
to the centroid shift at maximum amplification.
} \clearpage

\end{document}